\documentclass{elsart3}

\usepackage{amssymb}

\begin{document}

\begin{frontmatter}

\title{Novel surface anisotropy term in the FMR spectra of amorphous
microwires}

\author{M.W.~Gutowski\thanksref{IFPAN}\corauthref{corespn}}
\ead{gutow@ifpan.edu.pl}
\corauth[corespn]{Corresponding author. Tel.: (+48~22)-8436601-3122;
fax: (+48~22)-8430926.}
\author[IFPAN]{,~R.~\.Zuberek}
\author[CSIC]{, and A.~Zhukov}

\address[IFPAN]{Institute of Physics, Polish Academy of Sciences,
02--668 Warszawa, Poland}%
\address[CSIC]{Instituto de Ciencia de Materiales, CSIC, 28049
Cantoblanco, Madrid, Spain.}

\begin{abstract}
Some recent publications on ferromagnetic resonance in amorphous wires
mention presumably new kind of anisotropy, called there {\em
circumferential anisotropy\/}, as an explanation of various observed
spectral features.  In~this paper we argue that there is no special
reason to speak of the new kind of anisotropy, since the observed
spectra can be well described in terms of more traditional uniaxial and
surface anisotropies alone.
\end{abstract}

\end{frontmatter}

Our earlier investigations \cite{my} of FMR in magnetic, amorphous,
glass-coated microwires (diameter $\sim\!15~\mu$m, few mm long) have
shown, that no satisfactory description of the observed resonance lines
could be achieved with the usual form of the uniaxial anisotropy
energy, given by $E=K_{u}\sin^{2}\theta$, even if appended with an
extra term $K_{4}\sin^{4}\theta$.

The ferromagnetic wires are amorphous bodies, with no long range
chemical ordering, yet they cannot be considered as magnetically
isotropic objects. Very high cooling rate ($\sim\!10^{6}$~K/s) during
manufacturing process, together with the presence of a glassy,
non-magnetic outer shell are good reasons for considering the wire as
consisting of two essential parts: inner, cylindrically shaped inner
core and the outer coaxial tubular shell.  The interface between the
two is probably not very well defined, however it is obvious, that such
a~system cannot be treated in the same manner as uniform, classical,
free-standing `infinitely long cylinder'.  In this paper we will try
nevertheless to apply the classical theory of ferromagnetic resonance
in isotropic ellipsoids, assuming that the wire is magnetized
uniformly.

The spectra simulations \cite{my}, covering very broad ranges of the
searched anisotropy constants, up to $\pm 10^{7}$~erg/cm$^{3}$, were
carried out using methods of interval arithmetic \cite{intarith}, which
is extremely reliable and accurate technique. Neither the number of
observed re\-so\-nances, nor their positions could be reproduced
satisfactorily when only two uniaxial anisotropy terms,
$K_{u}$ and $K_{4}$, were taken into account. Being unsure what could
be the nature of the so called {\em helical\/} and {\em
circumferential\/} anisotropies \cite{oni}, we decided to study the
possible effects of the surface anisotropy on the re\-so\-nance
properties of cylindrically-shaped samples.  The assumption that the
surface anisotropy is of simple uniaxial type (easy or hard axis normal
to the wires' surface) leads to the contribution to the free energy in
the form $E\sim\!K_{s}\sin^{2}\theta$, indistinguishable experimentally
from the one describing the inner core. The uncontrolled stresses,
present at the wire-glass interface, and acting via inverse
magnetostrictive effect, due to symmetry reasons cannot be held
responsible for the orientation of easy/hard direction other than along
the wire.

Consider now two identical spins, treated here as classical vectors,
located at the circumference of a circle with radius $\rho$, and
separated by the distance $\delta$ one from another, both tangent to
the circle and lying in its plane.  The Heisenberg exchange energy of
their interaction is $E=-2J\left({\mathbf S}_{1}\cdot{\mathbf
S}_{2}\right)$, with $J$ denoting the exchange integral.  This energy
may be expressed as well as the function of the angle between both
spins: $E=-2JS^{2}\cos\phi_{12}$, where $S=\left| {\mathbf S}_{1}
\right|=\left| {\mathbf S}_{2} \right|$.  Given the spacing
$\delta$ and $\phi_{12} \approx \delta/\rho \approx\sin\phi_{12}$, we
can approximate
\begin{equation}\label{cosinus}
	\cos\phi_{12} = \sqrt{1-\sin^{2}\phi_{12}}
	\approx 1 - \frac{1}{2}\left(\frac{\delta}{\rho}\right)^{2}
\end{equation}
Therefore the difference of energies of this system when both spins are
lying on a straight line ($E_{1},\ \phi_{12}=0$) or are located on the
circle ($E_{2},\ \phi_{12}\ne 0$) is equal to $\left|E_{2}-E_{1}\right|
= \left|J\right|S^{2}\left(\frac{\delta}{\rho}\right)^{2}$,
i.e. it is inversely proportional to the squared radius of the circle. 
Exactly the same result would be obtained, if the spins were oriented
along the radii of the circle. We can identify this difference in
energy as arising entirely from the fact, that the surface, where the
spins are located, is curved.  In other words, the non-vanishing
surface anisotropy energy density is a geometric effect, proportional
to the squared local curvature of the magnetized surface and to the
strength of {\em isotropic\/} exchange interactions.  Perhaps this
effect should be called {\em exchange anisotropy\/}, since its origin
is similar to the one first described in \cite{exc}, with magnitude as
high as $7\times 10^6$~erg/cm$^{3}$ for $200$~\AA\ fine Co particles.

In order to calculate the surface contribution to the total anisotropy
one has to integrate proper expression along the elliptical path, with
continuously varying curvature, being the edge of the wires'
cross-section by the plane parallel to its magnetization vector.  Doing
so we obtain $E_{s}\propto\left|\cos\theta\right|$, therefore the
anisotropic part of the free energy density should read
\begin{equation}\label{full_energy}
	E= K_{u}\sin^{2}\theta + K_{4}\sin^{4}\theta
	+ K_{s}\left|\cos\theta\right|
\end{equation}
where $K_{u}$ is an ordinary uniaxial anisotropy term, modified by the
presence of the surface, $K_{4}$ is higher order uniaxial anisotropy
term, and $K_{s}$ is related to the following factors: intrinsic
surface anisotropy, the sample diameter and isotropic exchange
interactions.

The unusual form (\ref{full_energy}) of the free energy density derived
in this paper, suggests that the FMR spectrum of a~thin, glass coated,
amorphous wire can substantially differ from the one observed in
massive, free-standing samples.  In particular, there may be more than
just one resonance line for any given orientation of the external
magnetic field.  Additionally, the lack of smoothness of the anisotropy
energy at $\theta=\frac{\pi}{2}$ should be the reason for unusual shape
of the resonance absorption.  The peak, if any, observed in rather
strong external field, oriented nearly perpendicularly to the wire,
cannot be adequately described by the standard formalism of small
vibrations.  The free precession of a magnetization vector around its
equilibrium position will be, in this case, strongly disturbed (damped)
twice per every revolution.  This should lead to substantial increase
of the peak's linewidth.  Also, if the external field is not exactly
perpendicular to the wire, then the resonance peak should be clearly
asymmetric.  The overall shape of the resonance line should be
sensitive to the magnitude of the exciting field, i.e. to the applied
microwave power level.  All those effects were indeed observed
experimentally \cite{my}.


\begin{thebibliography}{9}

\bibitem{my} R.~\.Zuberek et al., phys. stal. sol.~(a), 196,
(2003), 205

\bibitem{intarith} R.B.~Kearfott, Euromath. Bull. 2, 95, 1996

\bibitem{oni} D.P.~Makhnovskiy et al., Phys. Rev. B 63 (2001),
144424

\bibitem{exc} W.H.~Meiklejohn, J. Appl. Phys. 33, 1328S, 1962

\end{thebibliography}
\end{document}